\documentclass[sigconf]{acmart}

\usepackage{booktabs} 

\setcopyright{none}

\usepackage{graphicx}
\usepackage{environ}
\usepackage{tikz}
\usetikzlibrary{calc,matrix}
\usepackage{cleveref}
\usepackage{float}
\usepackage[normalem]{ulem}
\useunder{\uline}{\ul}{}
\usepackage{subfloat}
\usepackage{graphicx}
\usepackage{amssymb}
\usepackage{xcolor}
\usepackage{paralist}
\usepackage{gensymb}
\usepackage{balance}

\usepackage{float}
\usepackage[normalem]{ulem}
\useunder{\uline}{\ul}{}
\usepackage{subfloat}
\usepackage{graphicx}
\usepackage{amssymb}
\usepackage{xcolor}
\usepackage{paralist}
\usepackage{gensymb}
\usepackage{booktabs} 
\usepackage{subfig}
\usepackage{booktabs}
\usepackage{multirow}
\usepackage[utf8]{inputenc}
\usepackage{multirow}
\usepackage{url}
\usepackage{makecell}

\usepackage{booktabs}
\usepackage{multirow}
\usepackage[utf8]{inputenc}
\usepackage{multirow}
\usepackage{url}
\usepackage{makecell}
\usepackage[colorinlistoftodos]{todonotes}

\newcommand{\specialcell}[2][c]{%
  \begin{tabular}[#1]{@{}c@{}}#2\end{tabular}}

\acmConference[RDSM'18]{the 2nd International Workshop on Rumours and Deception in Social Media}{October 2018}{Turin, Italy}
\acmYear{2018}
\copyrightyear{2018}

\begin{document}

\title[Characterizing the public perception of WhatsApp through the lens of media]{Characterizing the public perception of WhatsApp\\through the lens of media}

\author[J. Caetano et al.]{Josemar Alves Caetano\textsuperscript{1}, Gabriel Magno\textsuperscript{1}, Evandro Cunha\textsuperscript{1,2},\\Wagner Meira Jr.\textsuperscript{1}, Humberto T. Marques-Neto\textsuperscript{3}, Virgilio Almeida\textsuperscript{1,4}}
\affiliation{\textsuperscript{1} Dept. of Computer Science, Universidade Federal de Minas Gerais (UFMG), Brazil}
\affiliation{\textsuperscript{2} Leiden University Centre for Linguistics (LUCL), The Netherlands}
\affiliation{\textsuperscript{3} Dept. of Computer Science, Pontifícia Universidade Católica de Minas Gerais (PUC Minas), Brazil}
\affiliation{\textsuperscript{4} Berkman Klein Center for Internet \& Society, Harvard University, USA}
\email{{josemarcaetano, magno, evandrocunha, meira, virgilio}@dcc.ufmg.br, humberto@pucminas.br}

\begin{abstract}
WhatsApp is, as of 2018, a significant component of the global information and communication infrastructure, especially in developing countries. However, probably due to its strong end-to-end encryption, WhatsApp became an attractive place for the dissemination of misinformation, extremism and other forms of undesirable behavior. In this paper, we investigate the public perception of WhatsApp through the lens of media. We analyze two large datasets of news and show the kind of content that is being associated with WhatsApp in different regions of the world and over time.
Our analyses include the examination of named entities, general vocabulary and topics addressed in news articles that mention WhatsApp, as well as the polarity of these texts. Among other results, we demonstrate that the vocabulary and topics around the term ``whatsapp'' in the media have been changing over the years and in 2018 concentrate on matters related to misinformation, politics and criminal scams. More generally, our findings are useful to understand the impact that tools like WhatsApp play in the contemporary society and how they are seen by the communities themselves.

\end{abstract}

%
%



\maketitle

\section{Introduction}\label{sec:intro}
The messaging service WhatsApp is, as of 2018, one of the most rapidly growing components of the global information and communication infrastructure, counting with 1.5 billion users who send around 60 billion messages per day~\cite{15billion}.
This tool combines one-to-one, one-to-many and group communication by offering private chats, broadcasts and public group chats, through which users are able to send text and media (audio, image and video), as well as files in various formats.

According to data published by Statista~\cite{statista_shared}, more than half of the population of Saudi Arabia, Malaysia, Germany, Brazil, Mexico and Turkey were active WhatsApp users in 2017.
Also, the Reuters Institute Digital News Report 2018~\cite{reuters2018} shows a rise in the use of messaging applications, including WhatsApp, as sources of
news in several parts of the world. 
This report indicates that WhatsApp use for news has almost tripled since 2014 and it has surpassed Twitter as a communication system in many countries.
One of the alleged reasons for this is that users are looking for more private and secure spaces to communicate.
In addition to this, WhatsApp turned out to be an important platform for political propaganda and election campaigns, having held a central role 
in elections in Brazil, India~\cite{election_india}, Kenya, Malaysia, Mexico and Zimbabwe, for instance.
Also, WhatsApp has been frequently associated with the spread of misinformation and disinformation~\cite{theguardian_fears}.

Despite its prominence, continued growth and opacity, there has been an insufficient number of studies exploring the various aspects of WhatsApp and similar mobile messaging applications~\cite{Guo2018}.
Since WhatsApp provides encrypted end-to-end communication, it is a great challenge to conduct large-scale analyses on the behavior of its users. In this work, we take a different approach: instead of looking at inside the system, we focus on the public perception of WhatsApp from outside sources.
The goals of this paper are:
\begin{itemize}
\item to characterize how media in different countries interpret the role of WhatsApp in society;
\item to analyze the evolution of the perception of WhatsApp over time, from its creation 
until its massive popularization;
\item to comprehend how sensitive topics, such as politics, crime and extremism, are related to WhatsApp in different regions of the world and in distinct periods of time.
\end{itemize}
To achieve these goals, we explore different techniques: analysis of Web search behavior, co-occurring named entities and vocabulary, co-occurrence networks, topics addressed and textual polarity. According to our understanding, each of these methods is able to provide additional information about the perception of WhatsApp in the news articles investigated.
As a whole, our results indicate that the media has significantly changed its perception and portrayal of WhatsApp: while in the period before 2013 the focus of the news was on WhatsApp features, in the following years the tool started to be more associated with social issues, including the dissemination of misinformation.

This paper is organized as follows: in Section~\ref{sec:related_work}, we review a selection of works on WhatsApp and, more generally, on the use of textual datasets to understand social phenomena; in Section~\ref{sec:dataset}, we describe our methodology of data collection and the overall characterization of the datasets used in this investigation; next, in Section~\ref{sec:analysis_and_results}, we characterize the vocabulary, analyze the topics addressed and evaluate the polarity of the news articles contained in our datasets; finally, in Section~\ref{sec:conclusion}, we conclude the paper and present future directions of work.

\section{Related Work}\label{sec:related_work}

\paragraph{On the use of textual datasets to understand social phenomena}
Analyzing how a term is used over time and in a geographic location is important to help in the understanding of how cultural values, societal issues and customs are perceived by society and expressed through language~\cite{cambraia2013lexicologia,matore53}.
~\textit{Culturomics}, for example, is a concept proposed by~\citet{michel2011quantitative} referring to a method for the study of human behavior and cultural trends through quantitative analyses of texts, using sources like large collections of digitized books. 
Several studies explore this method to investigate topics such as the dynamics of birth and death of words~\cite{petersen2012statistical}, semantic change~\cite{gulordava2011distributional}, emotions in literary texts~\cite{acerbi2013expression} and characteristics of modern societies~\cite{roth2014fashionable}. Some works propose a complementary approach to~\textit{culturomics} by using historical news data~\cite{leetaru11}, analyzing European news media ~\cite{flaounas2010structure} or the writing style and gender bias of particular topics in large corpora of news articles~\cite{flaounas13}. Other works concentrate in specific events in history, such as the Fukushima nuclear disaster~\cite{lansdall2014coverage}, by using large datasets of media reports to understand aspects such as how the media polarity towards a topic changes over time.

Employing methods similar to the ones presented here,~\citet{cunha_socinfo18} investigate the perception and the conceptualization of the term ``fake news'' in the media, showing that contextual changes around this expression might be observed after the United States presidential election of 2016.
However, as far as we are concerned, this is the first work that uses these methods to examine in detail how the term ``whatsapp''
is being reported by news media in different parts of the world, making us able to analyze how important topics, such as misinformation, manipulation and extremism, might be associated with WhatsApp by societies.

\paragraph{On WhatsApp}
Despite the increasing use of WhatsApp in the world,
few studies about this instant messaging application are currently available.~\citet{DBLP:journals/corr/abs-1804-01473}
propose a data collection methodology for this application and perform a statistical exploration to indicate how data from WhatsApp public groups can be collected and analyzed. Also,~\citet{moreno2017whatsapp} collect WhatsApp messages to monitor critical events during Ghana's 2016 presidential election, and~\citet{Church:2013:WUW:2493190.2493225} analyze differences between WhatsApp and SMS messaging system using a large-scale survey.~\citet{7145326} investigate Facebook and WhatsApp traces collected from an European national wide mobile network and characterize the usage of both applications. The work of~\citet{seufert2016group} surveys users to investigate the usage of WhatsApp groups and, more specifically, its implications for mobile network traffic, while~\citet{DBLP:journals/corr/abs-1802-03393} collect personal information and 
messages from one hundred WhatsApp users
with the aim of 
understanding their usage patterns.

All of these works investigate a limited part of WhatsApp,
therefore offering a restricted understanding of how this application is used. Nevertheless, here we
study this tool using large datasets of external data provided by news articles containing the term ``whatsapp'' in different regions of the world
and covering the whole WhatsApp history, thus shedding light not exactly on its usage, but on how it is viewed from outside sources.

\section{Data Collection}\label{sec:dataset}

\begin{table*}[!ht]
\centering
\caption{(a) Number of news articles containing the term ``whatsapp'' in our NOW Corpus dataset according to the geographical origin of the corresponding news media; (b) Number of news articles containing the term ``whatsapp'' in both NOW Corpus and Brazilian news articles datasets according to the year of publication.}
 \begin{tabular}{l|l|r}
 \multicolumn{3}{@{}l}{(a) Geographical origin of news articles in our NOW Corpus dataset}\\
 \toprule
 \textbf{Region} & \textbf{Country} & \textbf{Occurrences} \\
 \midrule
 \multirow{3}{*}{\parbox{2cm}{The Americas}} & United States & 1,244 \\
    & Canada & 507 \\
    & Jamaica & 151 \\
 \midrule
 \multicolumn{3}{r}{Total: 5.73\% / 1,902}\\ 
 \midrule
 \multirow{4}{*}{\parbox{2cm}{Southeast Asia}} & Singapore & 2,889 \\
    & Malaysia & 2,578 \\
    & Philippines & 253 \\
    & Hong Kong & 124 \\
 \cmidrule{1-3}
 \multicolumn{3}{r}{Total: 17.61\% / 5,844}\\
 \midrule 
 \multirow{2}{*}{British Isles} & Great Britain & 2,251 \\
    & Ireland & 2,152 \\
 \midrule
 \multicolumn{3}{r}{Total: 13.27\% / 4,403}\\     
 \bottomrule
 \end{tabular}
 \quad
 \begin{tabular}{l|l|r}
 \toprule
 \textbf{Region} & \textbf{Country} & \textbf{Occurrences} \\
 \midrule
 \multirow{5}{*}{Africa} & South Africa & 5,274 \\
 	& Nigeria & 1,607 \\ 
 	& Kenya & 1,585 \\   
    & Ghana & 754 \\
    & Tanzania & 3 \\
 \midrule
 \multicolumn{3}{r}{Total: 27.79\% / 9,223}\\ 
 \midrule
 \multirow{2}{*}{Oceania} & Australia & 895 \\
    & New Zealand & 306 \\
 \midrule
 \multicolumn{3}{r}{Total: 3.62\% / 1,201}\\     
  \midrule
 \multirow{4}{*}{\parbox{2,8cm}{Indian subcontinent}} & India & 8,991 \\
    & Pakistan & 1,353 \\
    & Sri Lanka & 186 \\
    & Bangladesh & 82 \\    
 \midrule
 \multicolumn{3}{r}{Total: 31.98\% / 10,612}\\     
 \bottomrule
 \end{tabular}
 \vspace{1em}
 \newline
 \begin{tabular}{l|c|c|c|c|c|c|c|c|c|c}
 \multicolumn{10}{@{}l}{(b) Year of publication of news articles in both NOW Corpus and Brazilian news articles datasets}\\
 \toprule
 \textbf{Year} & 2010 & 2011 & 2012 & 2013 & 2014 & 2015 & 2016 & 2017 & 2018 & \textbf{Total}\\
 \midrule
 \textbf{Occurrences in NOW Corpus} & 4 & 41 & 145 & 393 & 1,101 & 1,642 & 7,266 & 11,677 & 14,636 & 33,185 \\
 \midrule
  \textbf{Occurrences in Brazilian news articles} &0 &0 & 4 & 91 & 427 & 785 & 904 & 888 & 948 & 4,047\\
 \bottomrule
 \end{tabular} 
 \label{tab:dataset}
\end{table*}

We use two large datasets of news articles
in this study. The first one is a collection of texts from the Corpus of News on the Web (NOW Corpus), which contains articles from online newspapers and magazines written in English in 20 different countries from 2010 to the present time~\cite{davies13now}.
This corpus is available for download and online exploration\footnote{\url{https://corpus.byu.edu/now/}} and, according to its author, it is, at the moment of our data collection, the largest corpus available in full-text format.
In 31 May 2018, we gathered all the news articles containing the 33,185 occurrences of the term ``whatsapp'' in the NOW Corpus. These news articles cover every year in the corpus (from 2010 to 2018) and comprise all 20 countries represented. These countries were then grouped into six regions based on their geographic locations (Africa, British Isles, Indian subcontinent, Oceania, Southeast Asia and the Americas).


Our second dataset includes articles collected from Brazilian online newspapers and magazines, all written in Portuguese, also containing the term ``whatsapp''. We searched for articles starting from 2010, but did not find any from 2010 and 2011 containing the term ``whatsapp'', so our second dataset contains news from 2012 to 2018. To build this dataset, we used the tool \texttt{Selenium}\footnote{\url{https://www.seleniumhq.org/}} to automate Web searches with the term ``whatsapp'' in the following ten major Brazilian news websites: \textit{Exame}, \textit{Folha de S. Paulo}, \textit{Gazeta do Povo}, \textit{G1}, \textit{O Estado de S. Paulo}, \textit{R7}, \textit{Terra}, \textit{Universo Online} (\textit{UOL}), \textit{Valor Econômico} and \textit{Veja}. The total number of occurrences of ``whatsapp'' extracted from these websites on 31 May 2018 is 4,047. Finally, we used the Python library \texttt{newspaper}\footnote{\url{https://pypi.org/project/newspaper/}} to collect the full texts of these news articles.

In \Cref{sec:coocurring_named_entities,sec:semantic_fiels,sec:coocurrence_networks,sec:topics_addressed,sec:polarity}, we analyze the news texts from the two previously described datasets. Table \ref{tab:dataset} shows the number of news containing the term ``whatsapp'' in our two datasets, according to the geographical origin of the corresponding news media and the year of publication of the news article.

In addition to these datasets, we also collected data from Google Trends\footnote{\url{https://trends.google.com/trends/}}, an online tool that indicates the frequency of particular terms in the total volume of searches in the Google Search engine. This tool also indicates the most common associated terms and the countries from which the highest volume of searches are originated from. It is also possible to filter these results for given periods. For our investigations, we collected data from searches made between 2010 and 2018, and use this information in Section \ref{sec:web_search_behavior}.

\section{Analyses and Results}\label{sec:analysis_and_results}
In this section, we discuss the outcomes of different analyses aimed to understand the perception of WhatsApp in the media. Each characterization is introduced by a description of how it may contribute to accomplish our goals, followed by the methodology employed and, finally, by a presentation and discussion of the results found.

\subsection{Web search behavior}\label{sec:web_search_behavior}
Before analyzing the public perception of WhatsApp through the lens of news articles from different regions of the world, we investigate whether it is possible to observe a change in the Web search behavior regarding the term ``whatsapp'' through time. We use data collected from Google Trends to perform this analysis.

\begin{figure}
\centering
  \includegraphics[scale=0.45]{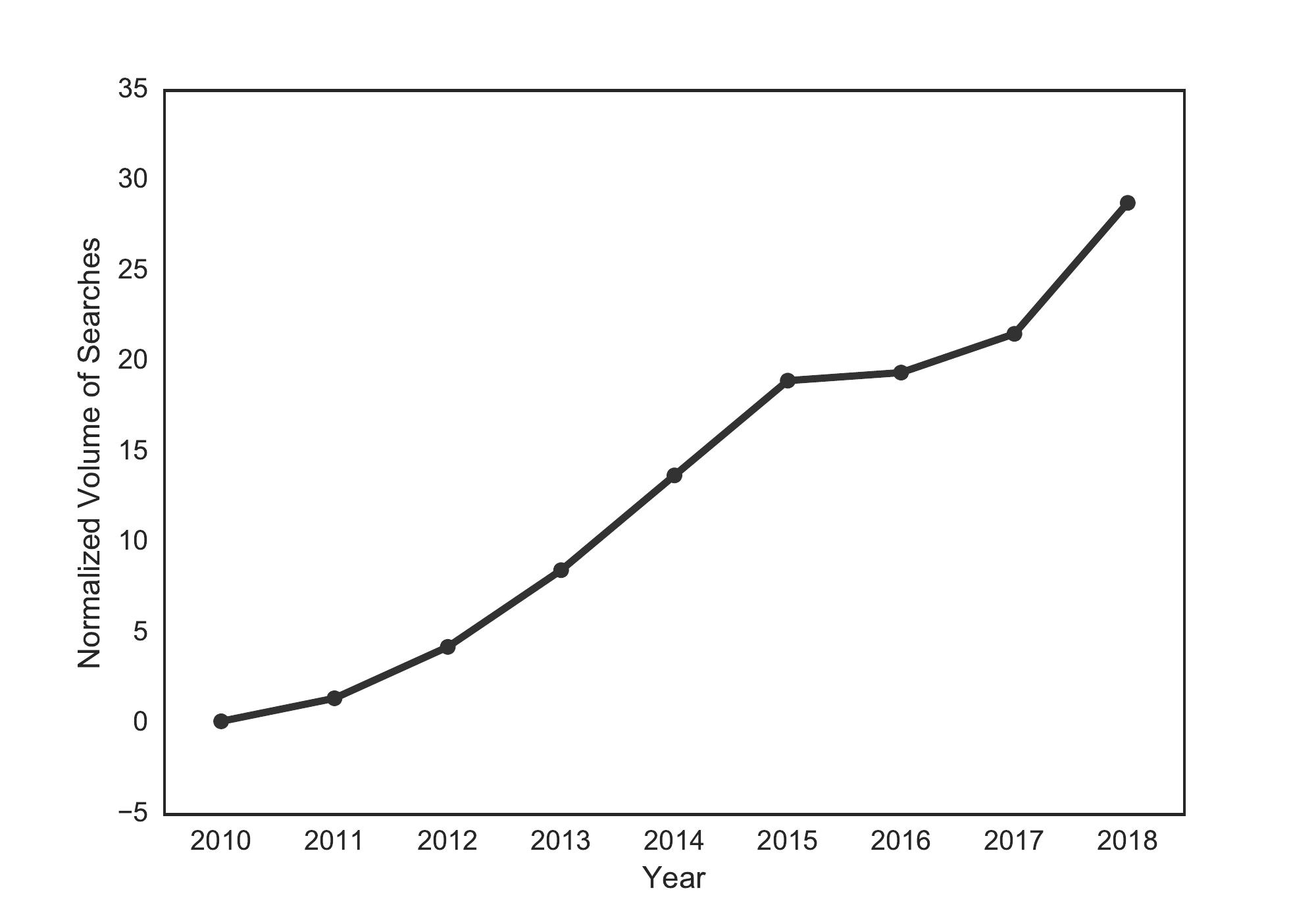}
  \caption{Normalized volume of queries for the term ``whatsapp'' in Google Search from 2010 to 2018}
  \label{fig:volume-searches-yearly}
\end{figure}

Our results show that, unsurprisingly, the number of queries on the Google Search engine for the term ``whatsapp'' is constantly growing since the release of this tool for Android devices in 2010, as indicated in Figure~\ref{fig:volume-searches-yearly}. 
Also, Table~\ref{tab:evolution-search} lists the five most frequent search terms employed by users who also searched for ``whatsapp'' from 2010 to 2018. 
Here, we notice a shift in the related terms through the years: in the first two years, most of the words are concerned with the download of the app (``download'', ``descargar'') and device compatibility (``blackberry'', ``iphone'', ``nokia''); then, from 2012 onwards, queries for ``whatsapp'' start to be linked to different topics, especially features of the tool (``status unavailable'', ``whatsapp encryption'', ``video status download''), but also content shared in WhatsApp (``imagens para whatsapp'', ``el negro del whatsapp'').

\begin{table}[h]
\centering
\caption{Most frequent queries related to ``whatsapp'' on Google Search per year}
\label{tab:evolution-search}
\begin{tabular}{c|c}
\toprule
\textbf{Year} & \textbf{Search terms} \\
\midrule
\multirow{1}{*}{2010} & blackberry, iphone, service, download, android\\
\midrule
\multirow{1}{*}{2011} & blackberry, nokia, app, download, descargar\\
\midrule
\multirow{2}{*}{2012} & error status, status unavailable,\\ & zello, sniffer download, double check\\
\midrule
\multirow{2}{*}{2013} & ios 7, imagens para whatsapp, intell app up,\\ & pagare whatsapp, baixa whatsapp\\
\midrule
\multirow{2}{*}{2014} & blaue haken, for nokia xl, masti.com, \\ & facebook compra whatsapp, blue ticks on whatsapp\\
\midrule
\multirow{2}{*}{2015} & whatsapp web, whatsapp reborn, caling feature, \\ & whatsapp transparante, llamadas whatsapp\\
\midrule
\multirow{2}{*}{2016} & negrita whatsapp, gb whatsapp, whatsapp encryption,\\ & el negro del whatsapp, cartas y whatsapp\\
\midrule
\multirow{2}{*}{2017} & video status download, whatsapp plus 2017, \\ &  status tamil, wasap weed, whatsapp storing\\
\midrule
\multirow{2}{*}{2018} & gb whatsapp 2018, plus 2018, \\ & call girls group link, browserling, whatsapp business\\
\bottomrule
\end{tabular}
\end{table}


\subsection{Co-occurring named entities}\label{sec:coocurring_named_entities}

In natural language processing, \textit{named entity recognition} is the task of extracting mentions of named entities -- that is, definite noun phrases referring to individuals, organizations, dates, locations -- in a text~\cite{bird2009natural}. We here extract the most mentioned named entities in our NOW Corpus dataset for each region and year of publication of the articles in order to understand who are the main actors related to the tool WhatsApp according to the media. In this paper, the co-occurrence is computed on a document level, so we consider all the entities that are mentioned in our news articles as co-occurring with the key-term ``whatsapp''.

To perform the named entity recognition, we use the Natural Language Toolkit (NLTK)\footnote{\url{http://www.nltk.org/}} classifier trained to recognize named entities. Since this tool does not support texts in Portuguese, we do not include the dataset containing the Brazilian news articles in this analysis.

Table~\ref{tab:entities-regions} lists the ten most mentioned entities in each different region considered in this investigation. Overall, we observe that the most mentioned entities accompanying the term ``whatsapp'' are usually other social media companies (``Facebook'', ``Twitter''), countries (``US'', ``India''), cities (``Dublin'', ``Delhi'') and demonyms (``African'', ``Australian''). When we analyze the continuation of the lists (not displayed here due to space constraints), we also find that US-American individuals like Mark Zuckerberg and Donald Trump are highly mentioned across the globe. However, local entities are also mentioned in their respective regions: among the entities not displayed in the table, the most mentioned persons or organized groups in each region are Mark Zuckerberg (the Americas and Oceania), Barisan Nasional (Southeast Asia), Paddy Jackson (British Isles), Uhuru Kenyatta (Africa) and Narendra Modi (Indian subcontinent). These findings suggest that news regarding the WhatsApp tool might deal with locally relevant entities -- which also are, most of the times, related to the local political scenarios.

\begin{table}[h]
\centering
\caption{Most mentioned named entities in each region (the entity ``whatsapp'' is excluded from the lists)}
\label{tab:entities-regions}
\begin{tabular}{c|c}
\toprule
\textbf{Region} & \textbf{Entities} \\
\midrule
\multirow{1}{*}{The Americas} & \specialcell[c]{Facebook, Google, US,\\ Twitter, Instagram, Apple,\\Android, American, Europe, China}\\
\midrule
\multirow{1}{*}{Southeast Asia} & \specialcell[c]{Facebook, Malaysia, India,\\Singapore, US, Malaysian,\\Google, Indian, China, Chinese}\\
\midrule
\multirow{1}{*}{British Isles} & \specialcell[c]{Facebook, Ireland, US,\\London, Irish, Google,\\British, Android, Dublin, Twitter}\\
\midrule
\multirow{1}{*}{Africa} & \specialcell[c]{Facebook, Twitter, South Africa,\\Nigeria, African, Kenya,\\Instagram, Africa, Nigerian, US}\\
\midrule
\multirow{1}{*}{Oceania} & \specialcell[c]{Facebook, US, Australia,\\Google, Australian, Apple, \\Instagram, Twitter,\\ Facebook Messenger, New Zealand}\\
\midrule
\multirow{1}{*}{Indian subcontinent} & \specialcell[c]{India, Facebook, Indian,\\Delhi, Mumbai, Pakistan,\\BJP, US, Twitter, Google}\\
\bottomrule
\end{tabular}
\end{table}

\begin{table}[h]
\centering
\caption{Most mentioned named entities in each year (the entity ``whatsapp'' is excluded from the lists)}
\label{tab:entities-year}
\begin{tabular}{c|c}
\toprule
\textbf{Year} & \textbf{Entities} \\
\midrule
\multirow{1}{*}{2010} & \specialcell[c]{Android, BlackBerry Messenger, BlackBerry,\\Kik, iPhone, Nokia, WiFi,\\Nokia N8, Symbian, India}\\
\midrule
\multirow{1}{*}{2011} & \specialcell[c]{BlackBerry, iPhone, Facebook,\\Android, Skype, SMS,\\Google, Nokia, Apple, US}\\
\midrule
\multirow{1}{*}{2012} & \specialcell[c]{Facebook, SMS, Android,\\India, iPhone, BlackBerry,\\US, Nokia, Skype, Twitter}\\
\midrule
\multirow{1}{*}{2013} & \specialcell[c]{Facebook, Android, Twitter,\\Google, Apple, India,\\Skype, SMS, BlackBerry, Indian}\\
\midrule
\multirow{1}{*}{2014} & \specialcell[c]{Facebook, Google, Twitter,\\US, Android, India, Apple,\\WeChat, China, Instagram}\\
\midrule
\multirow{1}{*}{2015} & \specialcell[c]{Facebook, India, Twitter,\\Google, US, South Africa, Android,\\Instagram, Skype, Apple}\\
\midrule
\multirow{1}{*}{2016} & \specialcell[c]{Facebook, India, Twitter,\\US, Google, Indian, Android,\\Instagram, Apple, iPhone}\\
\midrule
\multirow{1}{*}{2017} & \specialcell[c]{Facebook, India, Twitter,\\US, Indian, Instagram, Google,\\London, South Africa, China}\\
\midrule
\multirow{1}{*}{2018} & \specialcell[c]{Facebook, India, Twitter,\\US, Indian, Google, Instagram,\\Delhi, South African, Telegram}\\
\bottomrule
\end{tabular}
\end{table}

The ten most mentioned entities in each year are displayed in Table~\ref{tab:entities-year}. Among the entities that do not appear in the table due to space limitations, the most mentioned persons or organized groups in each year are: Steve Jobs (2011), Neil Papworth (2012), Mark Zuckerberg (2013 and 2014), Islamic State (2015 and 2016) and the Bharatiya Janata Party -- BJP (2017 and 2018). This indicates that, in general, the most relevant entities in the articles ceased to be linked to technology (Jobs, Papworth, Zuckerberg) and started to be related to social and political situations (Islamic State and BJP) from 2015 onwards, showing that WhatsApp gained importance outside of the world of technology and business.

\subsection{Semantic fields of the surrounding vocabulary}\label{sec:semantic_fiels}
Besides the analysis of the named entities that appear in the same news articles as the term ``whatsapp'', the investigation of the general vocabulary co-occurring with it is also valuable. One of the possible methods of performing such analysis is by observing the semantic fields (i.e. groups to which semantically related items belong)
of the words that appear in our news articles datasets, so to detect relevant concepts mentioned in the texts~\cite{cunha14_ht}.
Here, we use the tool \texttt{Empath}\footnote{\url{https://github.com/Ejhfast/empath-client}}~\cite{Fast:2016:EUT:2858036.2858535}, which provides a set of 194 lexical categories representing different semantic fields, each containing a list of words. Since Empath is available only in English, the dataset containing Brazilian articles was again not included in this analysis.

For this task, we first extracted all the words of each article and applied lemmatization -- that is, we grouped together their inflected forms so that they could be analyzed  as single items based on their dictionary forms (\textit{lemmas}). Lemmatization was performed employing the WordNet Lemmatizer function provided by the Natural Language
Toolkit and using verb as the part-of-speech argument for the lemmatization method, as in~\citet{cunha_socinfo18}.
Then, we counted the number of lemmatized words that appeared in each one of the Empath categories. In this phase, instead of using the absolute frequency of words, we normalized it by dividing the frequency of words in each category by the total number of categorized words.

\begin{figure}
\centering
  \includegraphics[scale=.55]{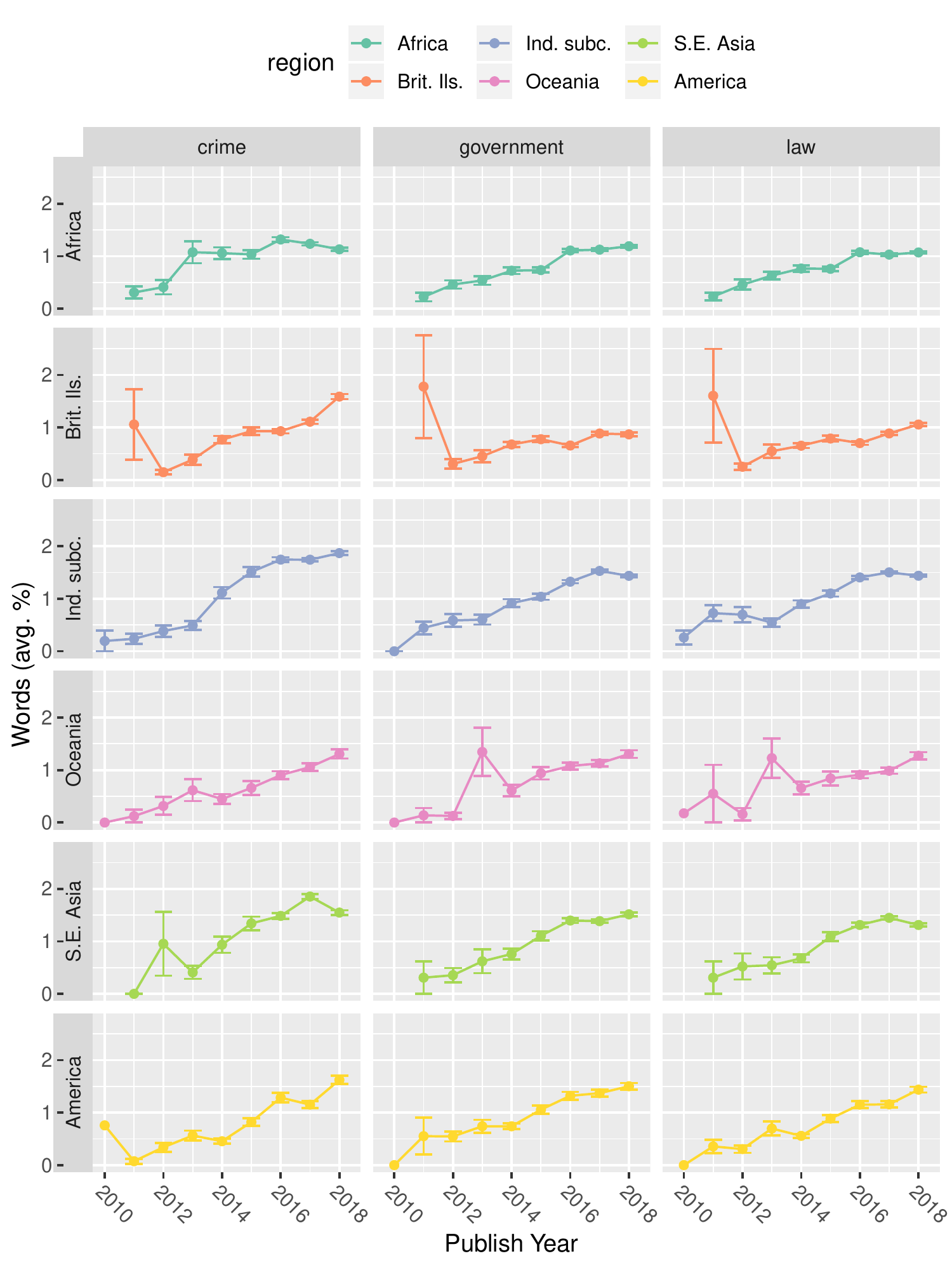}
  \caption{Average percentage of use of words from the semantic fields \textit{crime}, \textit{government} and \textit{law} in different regions and years (bars indicate standard errors of the mean values)}
  \label{fig:empath-avgbars}
\end{figure}

Since analyzing all the 194 Empath categories is impractical, we manually selected three relevant and noteworthy categories to scrutinize: \textit{crime}, \textit{government} and \textit{law}. In Figure~\ref{fig:empath-avgbars}, we present the average proportion of words belonging to these categories in news articles representing different regions across the years.
On the whole, we observe an overall increase in the proportion of words belonging to the three analyzed categories, with most of the peaks (such as the ones of 2013 in Oceania) probably due to political events (e.g. Australian federal election of 2013).
This finding indicates that words from the semantic fields \textit{crime}, \textit{government} and \textit{law} are being gradually more associated with WhatsApp in news from different regions of the world, corroborating the finding of Section \ref{sec:coocurring_named_entities} that shows an increase in the association of WhatsApp with social and political situations in recent years.

\subsection{Co-occurrence networks}\label{sec:coocurrence_networks}
Another possible analysis on the vocabulary accompanying a key-term in a corpus can be made through the observation of \textit{co-occurrence networks}. In our case, this method enables the visualization of the most relevant words that appear in the same news articles as the term ``whatsapp'' through the means of graphs. In this section, we consider both NOW Corpus and the Brazilian news articles dataset.

\begin{figure*}[h]
\subfloat[\label{fig:cooc-africa} Africa]{\includegraphics[scale=0.22]{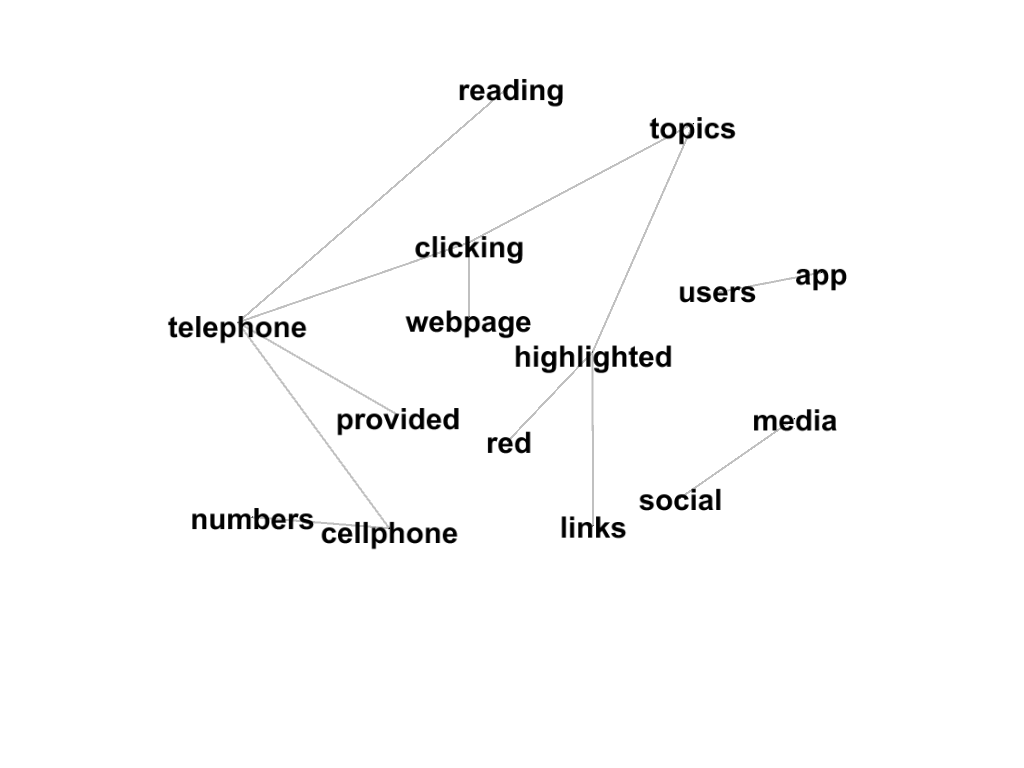}}
\quad
\subfloat[\label{fig:cooc-america} The Americas]{\includegraphics[scale=0.22]{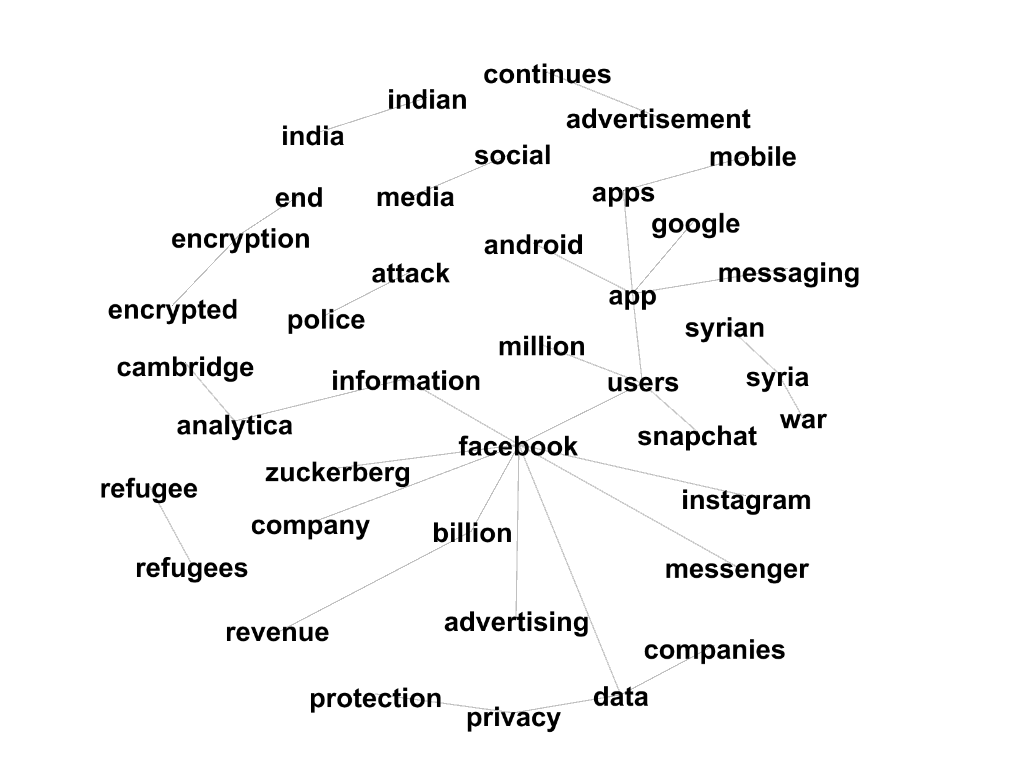}}
\\
\subfloat[\label{fig:cooc-britils} British Isles]{\includegraphics[scale=0.22]{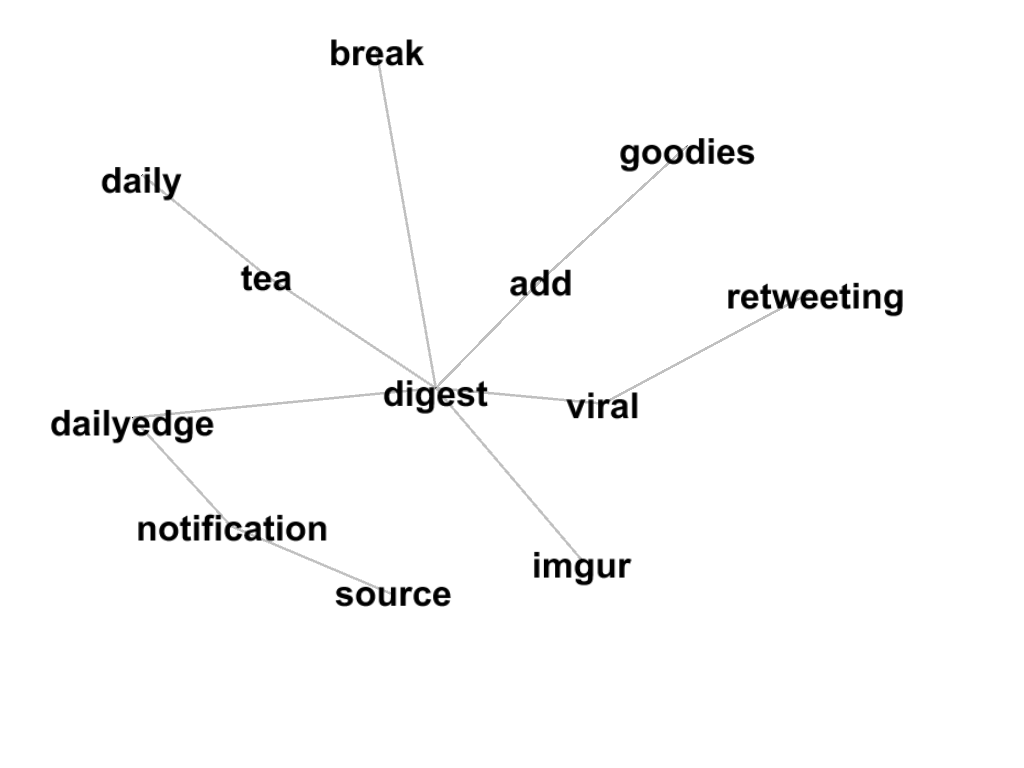}}
\quad
\subfloat[\label{fig:cooc-indsubc} Indian subcontinent]{\includegraphics[scale=0.22]{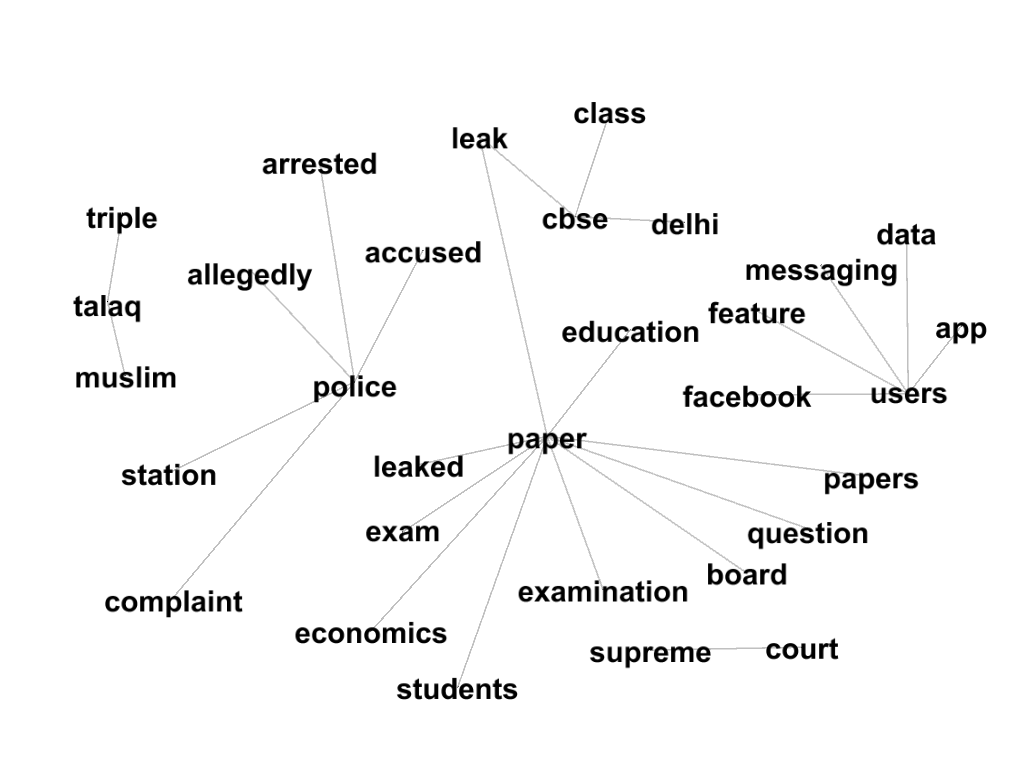}}
\\
\subfloat[\label{fig:cooc-oceania} Oceania]{\includegraphics[scale=0.22]{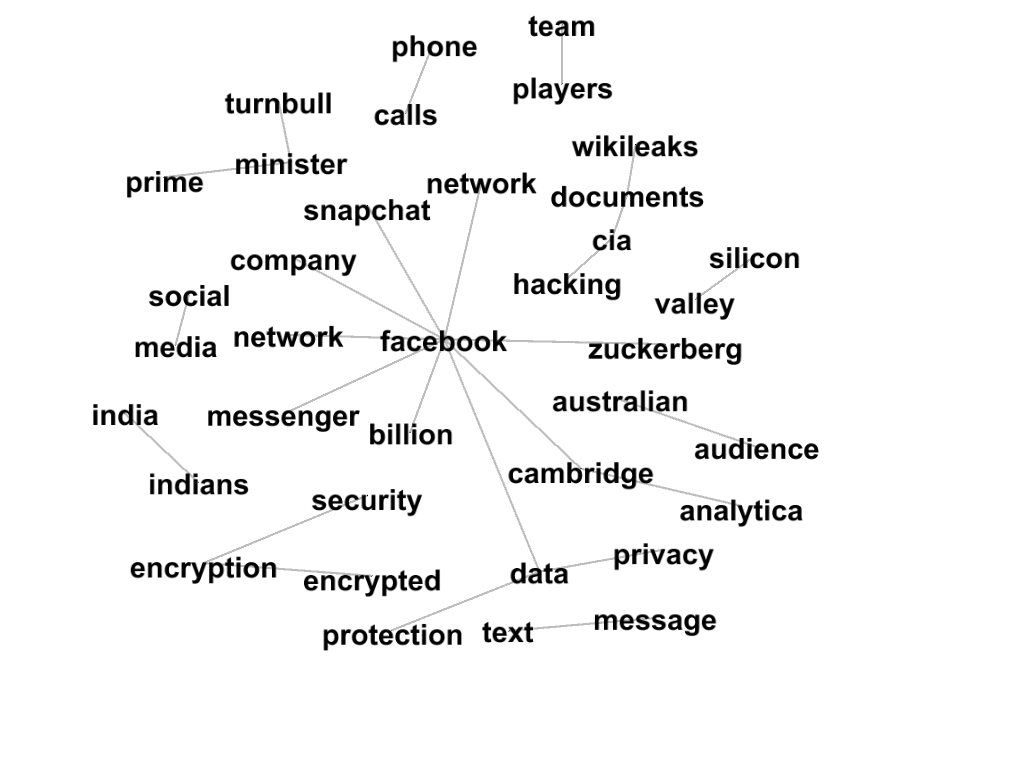}}
\quad
\subfloat[\label{fig:cooc-seasia} Southeast Asia]{\includegraphics[scale=0.22]{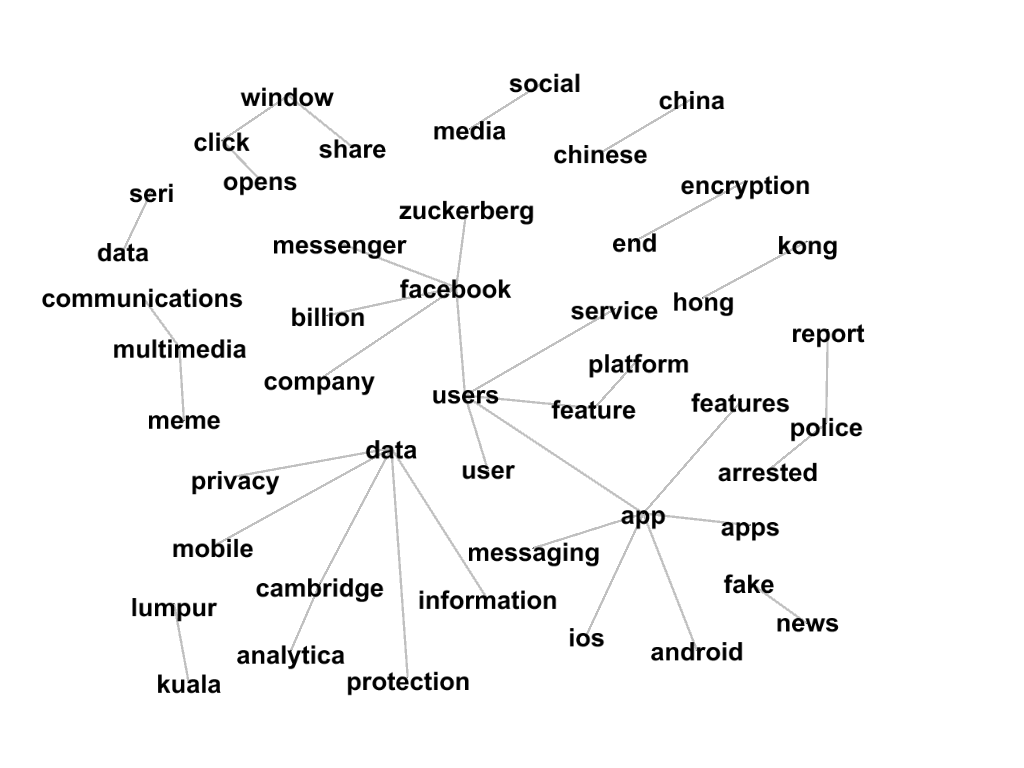}}
\caption{Co-occurrence networks for NOW Corpus news articles}
\label{fig:coocurrence-network-english}
\end{figure*}

For this analysis, 
we first extracted all the words from the articles and removed stop words using the lists provided by the Natural Language Toolkit for English and Portuguese.
Then, 
we extracted the most relevant words from each document by using the \textit{term frequency-inverse document frequency} (tf-idf) technique, that reflects how important a word is to a document in a corpus~\cite{rajaraman:2011}. We calculated the tf-idf for each pair $(document, word)$ and extracted from the document the top 50 words with the highest tf-idf scores.

In the following step, we counted the number of co-occurrences of the pairs of words. For each document, we obtained the list with its 50 most relevant words (according to tf-idf) and incremented by one the counter relative to each pair of words in this list (combination two by two). Instead of using the absolute count of articles in which two words co-occur, we normalized this value by dividing it by the total number of articles. At the end of this process, we obtained a graph in which vertices represent words and edges indicate their co-occurrence in the same texts. 

Since there is a considerable number of documents and news articles can be relatively long, the number of vertices and edges is large. For this reason, and due to the fact that our goal is to identify the most relevant relationships, we selected only the top 200 edges with the highest weights. Finally, we calculated the maximum spanning tree out of the remaining graph,
generating a graph that depicts the most relevant relationships in the format of a tree.

The final networks for the news written in English are presented in Figure \ref{fig:coocurrence-network-english} and clearly show some clusters
that generally represent different themes or specific events.
Some of the most relevant ones are: the 
``data'' clusters, related to privacy, regulation and data protection, containing words like 
``privacy'' and the name of information technology companies; 
the ``encryption'' clusters, related to the discussion towards WhatsApp's end-to-end encryption and containing words like ``security'' and  ``message''; 
and the ``crime'' clusters, with words like ``police'', ``attack'' and ``arrested''.

\begin{figure}[ht]
\centering
  \includegraphics[width=0.5\textwidth]{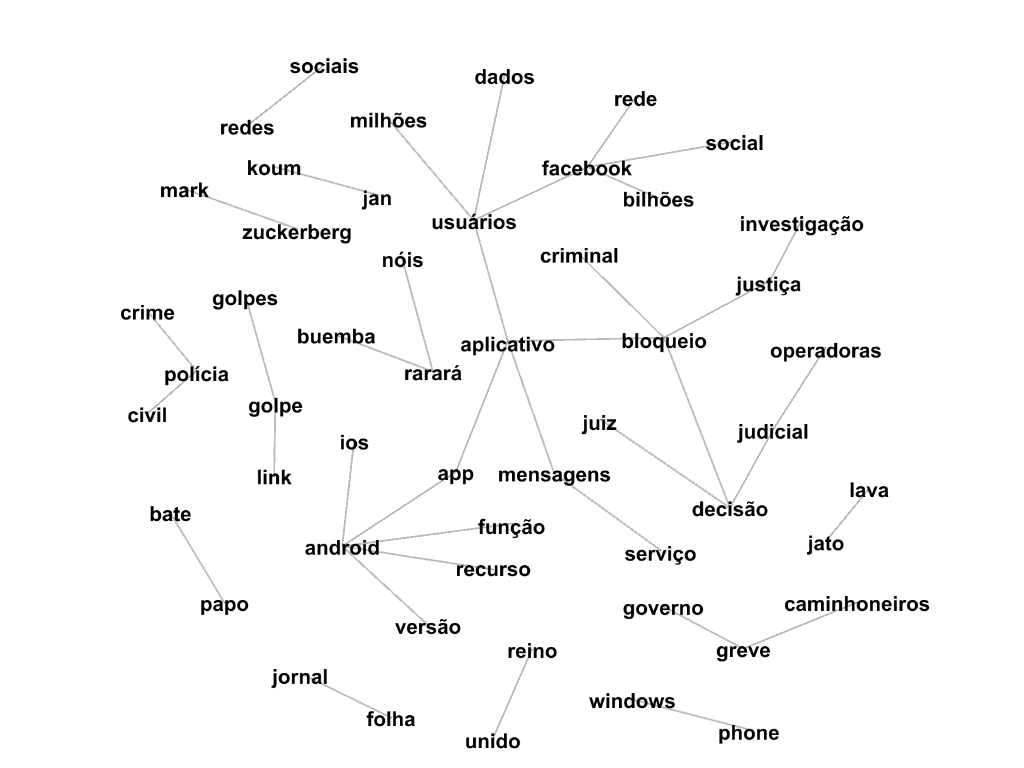}
  \caption{Co-ccurrence network for Brazilian news articles}
  \label{fig:coocurrence-network-brazil}
\end{figure}

The network regarding Brazilian news articles, presented in Figure \ref{fig:coocurrence-network-brazil}, also shows two of the aforementioned clusters: the ``data'' cluster (``dados'', ``usuários'', ``mensagens'') and the ``crime'' cluster (``polícia'', ``civil''). Besides that, it presents at least two other particularly interesting clusters. The first one is related to the government blocking WhatsApp in Brazil, with words like ``bloqueio'', ``justiça'' and ``operadoras''; and the second one is the ``truck drivers' strike'' cluster, represented by the words ``caminhoneiros'', ``greve'' and ``governo''.

\subsection{Topics addressed}\label{sec:topics_addressed}
In addition to investigate the vocabulary present in news articles mentioning WhatsApp, it is also possible to find the main topics addressed in the texts included in our datasets. We used \textit{latent Dirichlet allocation} (LDA)~\cite{blei2003latent} to automatically discover topics discussed in texts. For this task, we first lowercased and tokenized all the words in the datasets. Then, we removed stop words using, once again, the lists provided by the Natural Language Toolkit (after having added the word ``whatsapp'' to the lists, since it appears in all texts). Finally, we ran the LDA algorithm using the Python library \texttt{spaCy}\footnote{\url{https://spacy.io/}} for topic modeling. We used topic coherence score \cite{Newman:2010:AET:1857999.1858011} to choose the optimum number of topics \(k\) to be returned by the algorithm. For each region and year, the LDA model returned these \(k\) topics containing terms ordered by importance in the corresponding text. We then selected the most important topic as the representative of each region and year.

Table~\ref{tab:lda} shows the top-ranked ten terms produced by our LDA model representing the main topic for each region in each year.
Here, for the Brazilian articles, we translated the terms from Portuguese to English.

\begin{table*}[ht]
\small
\centering
\caption{Main topics for each year in each region}
\resizebox{\linewidth}{!}{%
 \begin{tabular}{c|c|c|c|c|c|c|c}
 \toprule 
\specialcell[c]{\textbf{Year}} & \textbf{The Americas} & \textbf{Southeast Asia} & \textbf{British Isles} & \textbf{Africa} & \textbf{Oceania} & \textbf{Indian subcontinent} & \textbf{Brazil}\\
  \midrule
2010 &\specialcell[c]{kik, service, federal,\\ user, message,\\ say, livingston,\\rim, blackberry,\\ growth, contact} &
\specialcell[c]{land, asli,\\ right, community,\\ customary, \\indigenous,\\ malaysian, \\ government,\\ recognition}
& --- & --- &
\specialcell[c]{app, phone, free,\\ message, text,\\ iphone, major,\\ blackberry,\\service, \\unlimited} &
\specialcell[c]{nokia, download, top,\\ smartphone, \\ mobile,\\ skype, america,\\ india, ovi, store} & ---\\
  \midrule
2011 & \specialcell[c]{business, small, \\wilton, owner, \\blackberry,patricio,\\ product, plan,\\ entepreneur, service} &
\specialcell[c]{free, text,\\ call, app,\\ viber, iphone,\\ network, phone, charge,\\ service} & 
\specialcell[c]{charge, dutch, \\net\_neutrality,\\ extra, mobile,\\ internet, law,\\ issue, state,\\ kpn, skype} & \specialcell[c]{mxit, knott\_craig, \\market, platform,\\ cheap, attention, \\buy, facebok,\\ mobile, smartphone} & \specialcell[c]{skype, iphone,\\ call, viber,\\ free, phone, \\mobile, platform,\\ android, text}  & 
\specialcell[c]{message, phone,\\ handset, android,\\ service,\\blackberry,\\ iphone, symbiam, \\text, app} & ---\\
  \midrule
2012 & \specialcell[c]{hutterite, wipf,\\ colony, website,\\ waldner,help, people,\\ life, medium,\\ social} & \specialcell[c]{hong\_kong, customer,\\ data, free, service,\\ messaging,\\ roam, hutchison, \\application, lead} &
\specialcell[c]{app, year, game,\\iphone, apple \\ communication, \\social, lync,\\ facebook, network} & \specialcell[c]{mobile, app, call,\\ nigerian, \\telecom, \\constitution,\\ growth, industry, \\ service, call, send} & \specialcell[c]{world, system, \\stop, late,\\ malware,\\ransomware,\\ sonicwall,\\ prevent,learn \\wannacry } & \specialcell[c]{phone, nokia,\\ asha, service,\\ price, launch,\\ company, internet,\\ cost, news} &
\specialcell[c]{client, message,\\ app, reconquer, \\telefonica,\\ europe, launch,\\ viber, intensify,\\ instantaneous} \\
  \midrule
2013 & \specialcell[c]{blackberry, bbm, \\user, company,\\ device, playbook,\\ service, app,\\ popularity, release} & \specialcell[c]{lau, speak, \\mobile, tencent, \\commercial,chat, \\ application, martin,\\ ltd, conference} & \specialcell[c]{screen, small,\\ blackberry, feel,\\ keyboard, camera,\\application,\\ distraction, fibre,\\ carbon, design} &  
\specialcell[c]{science, kelemu,\\agricultural, \\research, african,\\woman, school,\\ international, \\ develop, award} &
\specialcell[c]{privacy, user,\\ woman, data,\\ message, \\dutch, policy, \\server, agency,\\ address\_book} & \specialcell[c]{communal,\\ india, people,\\ indian, blood,\\muslim, riot, \\politician, \\secular, \\muzaffarnagar}& \specialcell[c]{market, technology, \\brazil, china, \\consume, \\ difficulty,\\ emerging, attempt,\\ domesticate,\\ invade} \\
  \midrule
2014 & \specialcell[c]{canada, refugee,\\ game, family,\\ libya, trip,\\ help, team, \\furniture, huddle} &
\specialcell[c]{government, right, \\party, state, \\political, law, \\country, leader,\\ election} & 
\specialcell[c]{party, johnson,\\ cohen, birmingham,\\ former, leader,\\britain, secretary,\\ prime\_minister, \\vote, election} & 
\specialcell[c]{show, music, event, \\african, art\\host, competition,\\ world, winner, black}  &
\specialcell[c]{minister,\\ party, leader,\\ government,\\ prime\_minister,\\ election,\\ national, senator,\\ republican,\\ president} & 
\specialcell[c]{pakistan, indian, \\terrorist, \\attack, muslim\\kill, freedom,\\ kashmir, army,\\ journalist}& \specialcell[c]{criminal, victim,\\ page, victim,\\ false, security, \\click, virus,\\ browser, federal} \\
\midrule
2015 & \specialcell[c]{jihadist, cabinet, \\edward, plausible,\\ outrage, \\nude, csec,\\ guido, chamber,\\ trove, inherently} &
\specialcell[c]{social\_media, job,\\ linkedin, online,\\ twitter, facebook,\\ company, professional, \\jobseeker, network} &
\specialcell[c]{terrorist, security, \\attack, intelligence, \\nisman, law, \\communication,\\ cameron, gchq, \\encryption} & 
\specialcell[c]{burundi, man, white,\\ election, protest, \\nkurunziza, president, \\police, party,\\ bujumbura} & \specialcell[c]{refugee, camp,\\ boat, data, web, \\australian,\\ service, phone,\\ use, communication} & \specialcell[c]{geeta, pakistan,\\ woman, girl,\\ ansar\_burney, \\karachi,\\ sushma\_swaraj,\\ police, comissioner,\\ raghavan}& 
\specialcell[c]{carriers,\\ application,\\ business, service,\\ block, justice, brazil, \\decision,\\ voice, president} \\
\midrule
2016 & \specialcell[c]{business, canada,\\ chera, border,\\ cbsa, trade, \\red\_tape, cfib, \\government, agency,\\ raise} & 
\specialcell[c]{hong\_kong, china,\\ chow, market,\\ hktdc, president,\\ wechat, product,\\ fair, party} & 
\specialcell[c]{scotland\_yard, \\religious, muslim,\\ country, british, \\authority, murder,\\ cafe, pope, bbc} & \specialcell[c]{nakuru, group, \\political, youth,\\ party, member,\\ jubilee, nyamira, \\governor, leadership} & \specialcell[c]{refugee, syrian,\\ aleppo, alkhuder, \\homescreen, \\ syria, earthquake,\\ greece, aircraft, \\assad} & \specialcell[c]{police, karachi,\\ punjab, ranger,\\ arrest, pakistan,\\ kashmir, kill,\\ medium, protest} & \specialcell[c]{government,\\ year, president,\\ fear, lula,\\ dilma, work, \\impeachment,\\ brazil, police}\\
\midrule
2017 & \specialcell[c]{market, report,\\ company, service,\\ user, include, help, \\facebook, mobile,\\ information} &
\specialcell[c]{group, lam, chat,\\ chat\_group,\\responsible, content,\\ campaign, china,\\ service, team} &
\specialcell[c]{attack, school, \\police, masood, \\westminster \\isis, terrorist,\\ arrest, kill,\\birmingham} & \specialcell[c]{president, nasa, \\kenyan, raila, \\leader, election,\\ political, party,\\ iebc, court} & \specialcell[c]{immigrant, \\country, \\immigration, \\trump, employee, \\visa, ban, refugee,\\ policy, president} &
\specialcell[c]{afghanistan,\\ kashmir, india, \\taliban, trump, \\policy, war,\\ obama, american,\\ troop} & \specialcell[c]{police, politics, \\government,\\ woman,\\ demonstration, \\ security, geddel,\\ prisoner, regime, \\arming} \\
\midrule
2018 & \specialcell[c]{meghan, harry, \\church, england,\\ ceremony \\wedding, \\prince, kate, \\ vow, royal\_wedding} & \specialcell[c]{facebook, zuckerberg,\\ data, scandal,\\ mistake, platform, \\prevent, ad,\\ authority, issue} & \specialcell[c]{facebook, people,\\ world, internet,\\ company,\\ technology, \\ social\_media, \\online,\\ change, new} & \specialcell[c]{student, school, \\university, high, \\education, parent,\\ water, health,\\ disease, study} & \specialcell[c]{national, party,\\ government, \\leader, state,\\support, michael, \\australian,\\ election, member} & \specialcell[c]{china, wechat,\\ tencent, \\ newsguard, app,\\chinese, company,\\ traffic, \\mall\_road,\\ authority}& \specialcell[c]{government,\\ president, \\truck\_drivers,\\ strike, minister, \\military\_coup,\\ group, support, \\deputy, world\_cup} \\
 \bottomrule
 \end{tabular}
 }
 \label{tab:lda}
\end{table*}

It is interesting to observe that,
between years 2010 and 2013, 
the main topic in almost all regions was related to WhatsApp features, device compatibility and differences between this application and other technologies, like
SMS.
In the Indian subcontinent, however, the main topic of 2013 was about riots and politics.

In the Americas, in years 2016-2017, the main topics of the news articles were also related to WhatsApp features. However, in 2014 and 2015 we can observe words like ``refugee'', ``libya'' and ``jihadist'', probably associated with events in the Arab world. In 2018, the main topic is related to the royal British wedding.

In Brazil, in 2014, we observe a topic shift to news related to criminal scams in WhatsApp. It is interesting to note that the main topic of 2015 is related to a Brazilian court decision to block WhatsApp in the whole country (because the company did not cooperate in a criminal investigation). 
In 2016, year of the impeachment of president Dilma Rousseff, the main topic contains words like 
``dilma'', ``impeachment'' and ``lula'', 
while the main topic in 2017 is also about politics, but containing more generic terms, such as ``politics'' and ``government''.
In 2018, however, we observe a clear dominance of terms related to Brazil truck drivers' strike, considered the biggest strike in the history of the country~\cite{truckers}.
In this occasion, WhatsApp played an essential role in the organization of the strike, differently from previous protests that were mostly coordinated through Facebook and Twitter.
This result reinforces the claims that
WhatsApp is a valuable tool to communicate and also to share political ideas in Brazil.

In the Indian subcontinent, we note that, between years 2013 and 2017, the main topics were related to political themes. In the year 2013, for example, words like ``riot'', ``muslim'' and ``muzaffarnagar'' are associated with the riots in Muzaffarnagar, 
when some rioters used WhatsApp to promote violence. In the years 2014-2016, rumors on terrorist attacks were disseminated through WhatsApp.
In 2017, the main topic seems to be associated to the decision of US president Donald Trump to not withdraw its troops from Afghanistan.

In Africa, 
in 2014, the words ``burundi'', ``election'' and ``protest'' are related to protests that occurred during the Burundian election. In this occasion, the government temporarily blocked messaging services, including Facebook, WhatsApp and Twitter~\cite{burundi}. In 2017, words like ``election'' and ``president'' are associated with the suspicion that disinformation and fake news were being used to influence Kenyans during the elections~\cite{kenya}.

There is also a clear dominance of words associated with terrorist attacks in 2015 and 2017 in the British Isles. These words are related to
the use of WhatsApp to organize these acts
~\cite{bbc_terrorism}. In Southeast Asia, news on WhatsApp are generally associated with comparisons with WeChat and, in the year 2018, news in this region were associated with the Facebook--Cambridge Analytica data scandal. In Oceania, in the years 2015-2017, the main topics were associated with refugees and immigration. WhatsApp played an important role during the Syrian Civil War in these years, since journalists and individuals living there used WhatsApp to communicate with people of foreign countries~\cite{syria}.

These results show that WhatsApp usage is highly associated with important political events in several regions of the world -- particularly in Africa, Brazil and India. The shift in the main topics addressed in the regions before 2013 (that were related to WhatsApp features and device compatibility) to, in the following years, political and criminal themes confirms results (presented in previous sections) that indicate a gradual increase in the association of this application with social and political situations.


\subsection{Polarity}\label{sec:polarity}
Our final investigation sheds light in another dimension of the news articles containing the term ``whatsapp'': now, we analyze the \textit{polarities} of the articles -- that is, whether the expressed opinions in the texts are mostly positive, negative or neutral. 
Here, we are interested in analyzing how the polarity of news articles related to WhatsApp changes over time and in different regions. 

To do this, we performed sentiment analysis
in each of the articles in our datasets using \texttt{SentiStrength}~\cite{Thelwall2010}, a tool that estimates the strength of positive and negative polarities in texts. This tool receives as input pieces of text and returns a score that varies from -4 (negative) to +4 (positive). 

\begin{figure}[ht]
\centering
  \includegraphics[scale=0.28]{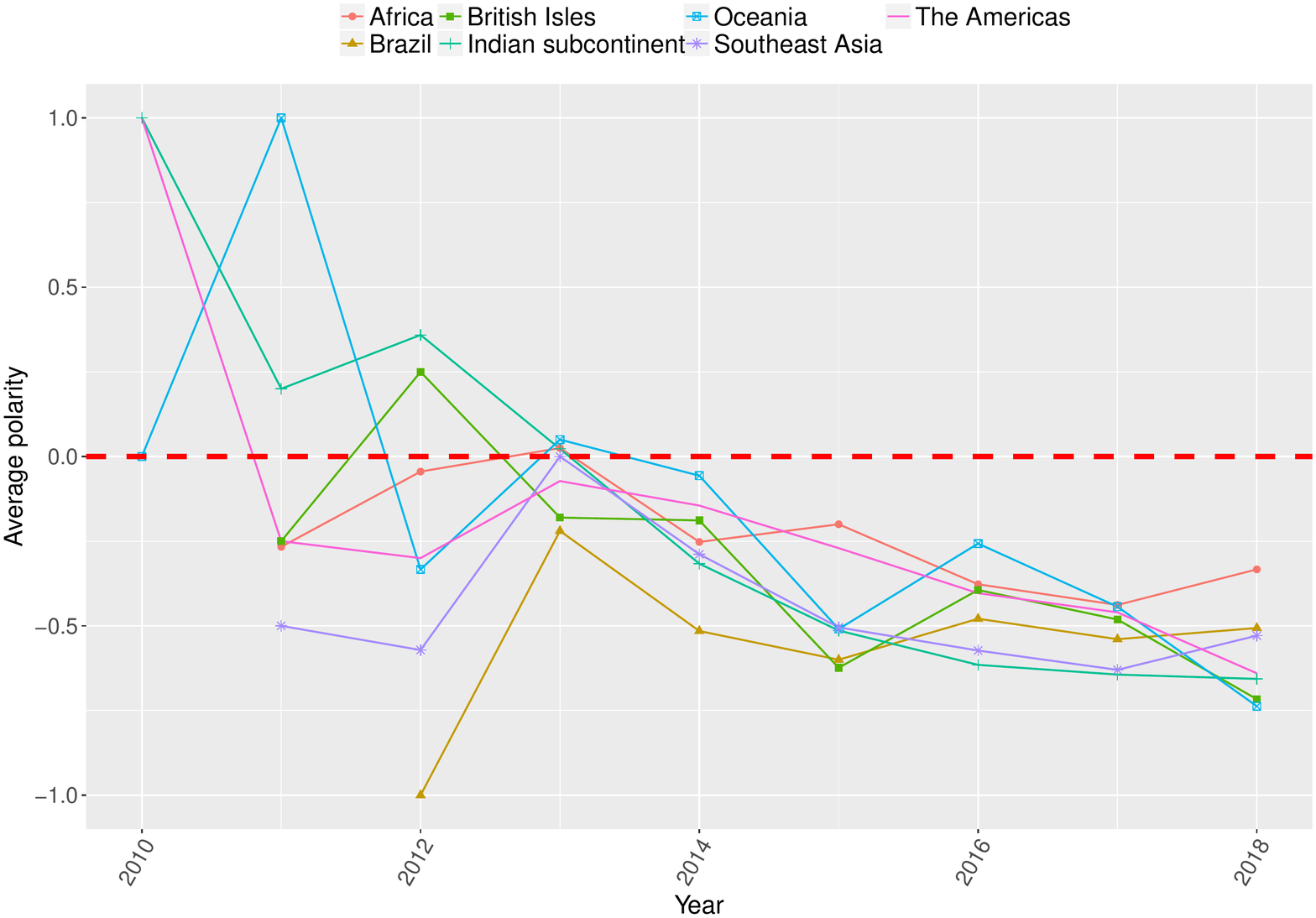}
  \caption{Average polarity of news articles from different regions containing the term ``whatsapp'' over time}
  \label{fig:all-regions-volume-news-country-yearly}
\end{figure}

\Cref{fig:all-regions-volume-news-country-yearly} depicts the average polarity of the news articles in each region and in each year, both in NOW Corpus and in the dataset of Brazilian articles.
We observe a major dominance of negative polarities in almost all regions and years, but especially after 2013. News articles containing the term ``whatsapp'' are becoming more negative over time probably because 
of the nature of the news articles themselves: in Africa, for instance, the term ``whatsapp'' occasionally appeared in news articles about refugees\footnote{\url{https://ewn.co.za/2017/12/11/un-seeks-1-300-places-for-refugees-from-libya}};
in India, in articles about the spread of fake news
that resulted in violence\footnote{\url{https://www.indiatoday.in/india/story/hyderabad-mob-of-200-pelts-stones-at\%2Dtransgenders-rumoured-to-be-child-lifters-kill-1-1242796-2018-05-27}}; 
in Southeast Asia and the Americas, 
in news
about the promotion of violence\footnote{\url{https://www.nytimes.com/2018/04/07/world/europe/london-murders-knife-attacks-stabbing.html}}; 
in Brazil, 
in news concerning criminal scams\footnote{\url{https://www1.folha.uol.com.br/tec/2016/02/1741888-golpe-que-promete-falsa\%2Dchamada-em-video-se-espalha-no-whatsapp.shtml}}.

\subsection{Summary of results}

The most relevant findings presented in this section can be summarized as follows:
\begin{itemize}
\item the interest for the term ``whatsapp'' is constantly increasing over the years,
as indicated by the rise of news about this tool and of Google Search queries for this term (Section \ref{sec:web_search_behavior});
\item this interest is being accompanied by a change of framing around the term ``whatsapp'' in the media -- from topics regarding WhatsApp features and technology to those related to misinformation, politics and criminal scams (Sections \ref{sec:web_search_behavior}, \ref{sec:semantic_fiels}, \ref{sec:coocurrence_networks}, \ref{sec:topics_addressed});
\item the polarity of news articles containing 
the term ``whatsapp''
is becoming more negative over time, probably due to the fact that this tool is being gradually more associated with
crimes, violence and fake news
(Section \ref{sec:polarity}).
\end{itemize}


\section{Concluding Remarks}\label{sec:conclusion}
In this paper, we present a quantitative analysis on the public perception of the messaging tool WhatsApp in news articles. For conducting our analyses, we used two datasets that cover the whole history of the application since its release for Android devices in 2010 until May 2018. The first of these datasets is a corpus of news articles written in English and published from 2010 to 2018 in 20 countries, while the second one contains Brazilian news articles published from 2012 to 2018. We also used data collected from Google Trends in one of our analyses.
 
Here, we investigated how media sources from different parts of the world have been reporting stories related to WhatsApp and whether the rise of the public interest in this application over time
was accompanied by changes on its perception by the media.
We observed changes in the vocabulary, in the mentioned entities, in the addressed topics and in the polarity of the articles mentioning the tool WhatsApp in our datasets.
In particular, we noticed a shift on media perception in almost all analyzed regions from the period before 2013 -- when the focus was on WhatsApp features and device compatibility -- to the following years -- when the application started to be gradually more associated with misinformation, manipulation and extremism, as well as with political and criminal activities.


The techniques and approaches proposed here can be used to measure the media perception of any company (or entity in general), but WhatsApp was chosen due to its influence in information (and misinformation) dispersion and to the fact that it has been related to topics such as extremism, corruption and political propaganda.
In future works, we intend to add more analyses, use news articles from others regions of the world where WhatsApp is popular (e.g. Germany, Indonesia, Malaysia) and compare the perception of WhatsApp in the media with the perception of it in other sources, like social networks and news articles comments. 
Also, we plan to compare the public perception of WhatsApp with the one of similar tools (e.g. Telegram, Facebook Messenger, WeChat) in order to understand which of them are more likely to be mentioned in certain types of news -- for instance, in political or crime-related news.

\bibliographystyle{ACM-Reference-Format}
\balance
\bibliography{bibliography}

\end{document}